\pdfoutput=1
\documentclass[12pt,preprint]{aastex}
\usepackage{graphicx}
\usepackage{natbib}
\begin{document}
\tighten

\title{A New Diagnostic of Magnetic Field Strengths in Radiatively-Cooled Shocks}

\author{
	P. Hartigan \altaffilmark{1},
        A. Wright \altaffilmark{1}
	}

\vspace{1.0cm}

\altaffiltext{1}{Rice University, Department of Physics and Astronomy, 6100 S. Main, Houston, TX 77005-1892} 

\begin{abstract}

We show that it is possible to measure Alfv\'enic Mach numbers, defined as the shock velocity in
the flow divided by the Alfv\'en velocity, for low-velocity (V$_{shock}$ $\lesssim$ 100 km$\,$s$^{-1}$)
radiative shocks. The method combines observations of bright forbidden lines with a measure
of the size of the cooling zone, the latter typically obtained from spatial separation between the Balmer
emission lines and the forbidden lines. Because magnetic fields become compressed as gas
in the postshock region cools, even relatively weak preshock magnetic fields
can be detected with this method. We derive analytical formulae that explain
how the spatial separations relate to emission-line ratios, and compute a large grid of radiatively-cooled
shock models to develop diagnostic diagrams that can be used to derive Alfv\'enic Mach numbers in flows.
Applying the method to existing data for a bright knot in the HH~111 jet, 
we obtain a relatively low Alfv\'enic Mach number of $\sim$ 2, indicative of a magnetized
jet that has super-magnetosonic velocity perturbations within it.  

\keywords{ISM: jets and outflows --- ISM: Herbig-Haro objects --- shock waves}

\end{abstract}

\section{Introduction}
\label{sec:intro}

Supersonic flows are common throughout the interstellar medium and govern
the dynamics of a variety of objects such as stellar jets \citep{frank14}
and supernovae remnants \citep[e.g.][]{reynolds08},
and also play an important role in affecting the observed spectra in
protoplanetary nebulae \citep{riera03} and novae shells \citep{williams13}.
In cases where shocked gas cools by radiating emission lines,
line ratios provide measurements of the densities, temperatures,
ionization fractions, and abundances throughout the flow \citep[e.g.][]{laz06}. Emission line
widths are a powerful way to quantify the amount of turbulence and nonthermal 
motions present in these objects, and high-resolution images that detect proper motion have 
begun to open up the time-domain, making it possible to observe shear, turbulent
mixing, fluid-dynamical instabilities and cooling in real time \citep{hartigan11}.

This wealth of information not only defines the physical characteristics
at each location in flows and helps to clarify the nature of the
driving sources, but can also identify the most important atomic
or plasma processes that operate in supersonic flows.
For example, the presence of both a narrow and a broad velocity component
in the H$\alpha$ emission at the blast wave of many supernovae remnants 
arises from the high cross section of charge exchange 
between postshock ions and neutrals \citep{chev80}. The resulting H$\alpha$
emission-line shapes constrain the velocity and orientation of the shock front,
and the degree to which ion and electron temperatures are equilibrated
behind the shock \citep{ghav01}. The relatively large line width associated with
the narrow component points to the possibility of a cosmic-ray precursor \citep{raymond11}.
In the case of stellar jets, H~II regions visible ahead of the
strongest shocks define the regime where radiative precursors begin to affect
shock dynamics, and the presence of transient bright spots where shock waves intersect
leads to the physics of Mach stem formation and evolution \citep{hartigan11,frank14}.

Although magnetic
fields can play an important role in shocked flows on both large and small scales,
especially for systems like stellar jets where the driving source is likely to be an
MHD disk wind of some sort, it is difficult to infer field strengths in
stellar jets directly from optical emission line spectra. Optical lines show no discernible Zeeman splitting, 
and emission line ratios in shocked gas are degenerate with respect to field strength in the sense that
spectra from low-velocity shocks without a preshock magnetic field closely resemble those
from a higher-density, higher-shock velocity object with a field \citep[Table 7 of ][]{hmr94}.
If a bow shock is well-resolved spatially and emits in [O~III]
near its apex, one can use the extent of the [O~III] emission and the shape
of the bow to infer the shock velocity, and the H$\alpha$ flux to determine 
the preshock density. With this information in hand, the observed electron densities from
an emission line ratio like [S~II] $\lambda$6716/$\lambda$6731 
give the compression, and hence the strength of the preshock and the postshock
field that is oriented in the plane of the shock 
\citep[i.e., perpendicular to the normal of the shock front;][]{morse92,morse93a}.

Most jets do not emit detectable synchrotron emission, but nonthermal radio continua have
been measured in a few bright knots, which are typically unresolved spatially. 
Fields have also been inferred from synchrotron continuum flux levels and from
polarization measurements in a few cases \citep{ray97,chrys07,cg10}. However, converting
these measurements to an estimate of the field strength 
requires knowledge of the electron density in the radio-emitting region immediately
behind the shock where the density is changing rapidly, and also involves
assumptions about the filling factor of the gas, the degree of ionization, and
the power spectrum of the electrons.

Magnetic field strengths inferred from the resolved-bow-shock method and from
continuum polarization both indicate surprisingly weak fields.  \citet{morse92,morse93a}
found magnetic fields of $\sim$ 30$\mu$G in the preshock gas ahead of bright bow shocks
in HH~34 and HH~111. 
The observed compression ratio between the preshock gas and the radiating gas is $\sim$ 40
(a factor of $\sim$ 4 at the shock and another $\sim$ 10 in 
the cooling zone), so that the field strength is $\sim$ 1mG where forbidden lines emit
at temperatures of $\sim$ 7000K. The electron densities and ionization fractions
determined at that location in the postshock gas
imply a total density of $\sim$ $3\times 10^4$ cm$^{-3}$, resulting in an Alfv\'en speed of only
$\sim$ 10 km$\,$s$^{-1}$ in the dense cooling zone. Hence, while the magnetic signal speed is
comparable to the thermal velocity within shocked knots in the flow, it is 
negligible compared with the bulk flow velocities of $\sim$ 300 km$\,$s$^{-1}$.
Field strengths inferred from radio continuum observations of
the unusually strong shocks in HH~81 are also only 100$\mu$G \citep{chrys07}.
These results are in agreement with proper motions of jets like HH~111 (oriented
in the plane of the sky), where internal velocity variations on the
order of 40 km$\,$s$^{-1}$ produce shocks and not magnetic waves, implying that
the magnetic signal speeds in the jet are at most a few tens of km$\,$s$^{-1}$
\citep{hartigan01}.

Two factors help to account for the anomalously low fields observed in stellar jets.
First, a pulsed flow with supersonic velocity perturbations will collect magnetic field
into knots, while rarefactions in the flow stretch the field out.
This effect lowers the magnetic signal speed ahead of jet knots and allows smaller
velocity perturbations to form shocks that otherwise would not do so \citep{hartigan07}.
Another way to reduce the field in the jet is to remove it by reconnection. Some evidence exists
for reheating at tens of AU from the sources both from high-spatial resolution observations of
optical emission lines made with HST \citep{hm07}, and X-ray observations
of quasi-stationary knots that are offset spatially from their sources \citep{xray}.
In laboratory settings, magnetic towers become kink-unstable close to their driving
sources \citep{moser12}, and the same thing happens in some numerical simulations of disk winds
\citep{staff15}. Kink instabilities could result in reconnection as the field geometry
becomes increasingly complex.

With the importance of measuring field strengths in MHD flows self-evident, it is 
worthwhile to revisit the subject to see whether it may be possible to improve upon optical
emission-line analysis of jets. Optical emission lines make the best connection
to the source of any MHD wind, because they arise within the material driven directly
from the source.  In contrast, molecular emission lines in these outflows are
typically slower and wider, and often come from a sheath of material entrained by the jet.
It has been known for almost two decades that H$\alpha$ and [S~II]
emit from different locations in jet shocks, with H$\alpha$ coming from
collisional excitation immediately behind the shock and [S~II] following in a 
cooling zone \citep{heathcote96}. Surprisingly, this observational constraint
has not been used to help infer field strengths, despite the fact that even a weak
preshock magnetic field affects this spatial offset in a significant way
\citep[e.g. compare models B100 and E100 in ][]{hrh87}.

In this work we show that {\it the additional constraint of a spatially-resolved
cooling-zone breaks the degeneracy of the line ratio information and makes
it possible to estimate Alfv\'enic Mach numbers of radiative shock waves stellar
jets directly, using only quantities that are readily observable}.  We discuss the 
physics of magnetized cooling zones in Sec.~\ref{sec:physics}, and present
a large grid of shock models and their resulting diagnostics in Sec.~\ref{sec:models}.
Although the ideal observations have yet to be made, we collect the best
available data to obtain preliminary estimates of the Alfv\'enic Mach number for 
a shock in the HH~111 jet
in Sec.~\ref{sec:application}, and conclude with a short Summary section.

\section{Measuring Alfv\'enic Mach Numbers Using Emission Line Ratios and Cooling Distances} 

\subsection{Physics of Magnetized Cooling Zones}
\label{sec:physics}

As gas enters a shock front, the pressure, temperature, and density immediately
behind the shock are determined by the Rankine-Hugoniot equations, derived by
balancing the mass, momentum, and energy fluxes on either side of the shock
\citep[e.g.][]{shockref}. Material that has passed through the shock is heated
as the bulk kinetic energy in the flow transforms into thermal energy. The hot gas
then cools by emitting line- and continuum-radiation, and within this cooling
zone the temperature decreases and the density increases in response to the energy
lost by radiation.  The energy equation in the cooling zone of a partially-ionized
steady-state shock is easily described by a formalism that relates the
enthalpy per nucleus in the gas to the radiative energy loss rate \citep{raymond79}.
Fig.~\ref{fig:non-B-case} depicts an
example of the spatial distribution of emission-lines in the cooling zone of a
typical radiative shock front.

Several types of magnetized shocks are possible depending on the orientation
of the field relative to the shock front \citep[e.g. Chapter 7 of][]{gb05}. 
In general, one must solve a cubic equation known as the shock adiabatic to obtain
the jump conditions for density and pressure. 
If the field is oriented perpendicular to the shock normal (perpendicular to the velocity vector
and parallel to the plane of the shock), there is a single real solution. In this case, an
MHD shock forms as long as the incident velocity exceeds that of the fast magnetosonic speed
V$_{fms}$ = [V$_A^2$ + C$_S^2$]$^{1/2}$, where C$_S$ is the sound speed in the preshock gas and
V$_A$ is the Alfv\'en speed in the preshock gas. 
If the direction of the field is aligned parallel to the normal to the plane of
the shock (i.e., along the velocity vector) two solutions are possible, one where
the jump conditions are the same as those of the B=0 hydrodynamical case, and another
so-called switch-on shock where surface currents impart a tangential component to the field
in the postshock gas. 

For the general case where the field is at an angle to the surface,
one can transform into the deHoffmann-Teller frame where {\bf v} $\parallel$ {\bf B},
and the equations have slow, intermediate, and fast solutions that correspond
as the compression approaches unity to,
respectively, the slow MHD wave speed, the Alfv\'en speed, and the fast MHD speed. 
Once the Alfv\'enic Mach number of the flow $\gtrsim$ 2, only the fast solution remains.
Throughout this paper we have used the fast MHD solution, and 
when we refer to a magnetic field strength we mean the component perpendicular to the
flow that lies in the plane of the shock.
This component should be the dominant one for a typical jet geometry because
in an MHD disk wind at locations far from the source, 
fields in jets should be mostly toroidal and therefore lie in the plane of the shock for 
a velocity-variable flow unless turbulence or kink instabilities randomize the direction.

The addition of magnetic fields alters the structure of the cooling zone in important ways. 
Because even small ionization fractions ($\gtrsim$ 10$^{-4}$) suffice to
tie the magnetic field to the gas, the magnetic field strength B is proportional
to the gas density n throughout the entire cooling zone. The total pressure
in the cooling zone is roughly constant, so n $\sim$ T$^{-1}$ as long as thermal
motions dominate the pressure.  However, the
magnetic pressure B$^2$/8$\pi$ $\sim$ T$^{-2}$, so as
the temperature in the cooling zone drops, the ratio of the magnetic pressure
to the thermal pressure rises, and typically exceeds unity in the forbidden-line zone 
(see Figs.~\ref{fig:B-case},~\ref{fig:B-case-zoom}).
Thus, even if the preshock magnetic pressure is orders of magnitude below the
ram pressure at the shock front, by the time the flow radiates forbidden lines, the
field usually dominates the pressure. A strong field in the cooling zone affects
the observed line radiation mainly by (i) lowering the density, and (ii) by
expanding the size of the cooling zone. 

From an observational standpoint, one can easily measure the electron 
density in the cooling zone from, for example, the
red [S~II] ratio $\lambda$6716/$\lambda$6731, but this measurement
alone does not constrain the field strength because the density in the cooling zone
also depends linearly upon the preshock density. The
shock velocity also matters because higher shock velocities produce hotter postshock temperatures,
which result in a larger temperature drop and subsequent larger density
increase once the gas reaches the forbidden line zone. In addition, the preshock ionization
fraction of the gas also has a small effect on cooling zone densities
because more neutral preshock conditions result in lower postshock temperatures
as some of the ram energy goes into ionizing the postshock gas \citep{cr85}.  In practice, many
emission lines occur in the cooling zone, and combining all of them narrows
the possible parameters considerably. However, as noted in the Introduction,
one cannot infer magnetic field strengths from emission line ratios alone
because adding a preshock field affects emission line ratios in a similar manner to
lowering the preshock density and shock velocity.

The second main observational effect of a weak preshock field - an expanded cooling zone
distance - holds some promise as a diagnostic, and
has not been explored completely from a theoretical or observational
standpoint. In non-magnetic shocks, this
distance scales approximately as a power law, V$_S^x$n$_\circ^{-1}$, where V$_S$ is the
shock velocity, n$_\circ$ is the preshock density, and the power $x$ $\sim$ 4.0 $-$ 4.5 \citep{hrh87}.
Observationally, one can only measure the cooling zone size if there is an
emission line that peaks at or near the location of the shock, another that
peaks within the cooling zone, and this separation is resolvable with current instrumentation.
Shock velocities in stellar jets are well below the threshold where
radiation ionizes the preshock gas ($\sim$ 100 km$\,$s$^{-1}$), so
a significant fraction of the H that enters the shock is neutral. 
This neutral H becomes excited by collisions at the shock and radiates strongly in
Balmer emission lines in a process similar to that which produces broad and narrow
Balmer filaments in some non-radiative supernova remnants \citep{snrref}.
Hubble images of stellar jets show that H$\alpha$ follows sharp arc-shaped features that
define the locations of the shocks, while emission lines such as [S~II] $\lambda$6731
follow behind these shocks with an observable spatial separation \citep{heathcote96}.
Recent images of HH~2 show enhanced Balmer decrements (H$\alpha$/H$\beta$) at the shock
fronts, as expected for collisional excitation of hydrogen \citep{raga15}.

Cooling zone sizes become more difficult to measure when the shock velocity is high
($\gtrsim$ 100 km$\,$s$^{-1}$), because the nature of the
Balmer emission changes at higher shock velocities. 
Postshock temperatures in high-velocity shocks exceed $10^6$K, and the many
different species of highly-ionized metals present at these temperatures emit energetic
photons capable of ionizing H.  As a result, most of the Balmer emission no longer occurs from
collisional excitations of neutral H immediately behind the front, 
but instead originates from recombination at a secondary peak in the cooling zone
where Lyman continuum photons emitted by the hot gas near the front 
are absorbed.  In these high-velocity shocks,
radiation also affects the forbidden line fluxes as electrons ejected by
the Lyman continuum photons excite the low-energy states of metals by collisions.
While full radiative transport is included in the models we present in Section~\ref{sec:models}, 
radiation plays almost no role in the results of this paper
because, as noted above, the internal shock velocities 
within stellar jets derived from both emission line ratios and
differential proper-motion measurements are typically $\lesssim$ 80 km$\,$s$^{-1}$
\cite{hmr94}. 

Our models show that this secondary H$\alpha$ peak \citep[e.g.][]{rb91} spreads out over the
same temperature range as the forbidden lines of [N~II] and [S~II], and its integrated flux is
typically less than that of the integrated flux of the H$\alpha$ at the front as long as
the preshock ionization fraction X$_H$ $\lesssim$ 0.5. In practice, this second peak should
not confuse the interpretation for most stellar jets
because it lies on top of the forbidden line emission
and does not appear as a distinct sharp feature. Once shock velocities exceed
$\sim$ 100 km$\,$s$^{-1}$, and the preshock radiation field becomes intense enough to
fully preionize the incoming gas, it will become impossible to measure
cooling zone distances from H$\alpha$. However, in those cases one could in principle
look for high-ionization lines such as those from N~V, O~VI, and C~IV, or even X-ray emission
to define the spatial location of the shock front and then measure this offset relative
to the forbidden-line cooling zone. 

Within the cooling zone there are multiple possibilities for emission-line
tracers.  For example, the lines of [O~I] and [S~II] both 
peak around 7000K, while those from [N~II] emit on average a bit closer to the front, where
T $\sim$ 10$^4$K.  In this paper we define a cooling zone distance d$_C$ as the 
average distance from the shock of photons from the red [S~II] lines $\lambda\lambda$6716+6731.
In practice, this is effectively equal to the spatial offset between H$\alpha$ and
[S~II] because H$\alpha$ peaks immediately behind the shock.
As we show in the next section, this measurement 
of d$_C$, combined with the [S~II] $\lambda$6716/$\lambda$6731 flux ratio,
makes it possible to infer Alfv\'enic Mach numbers in low-velocity radiative shocks. 
This method is independent of abundances and reddening, and uses the brightest 
emission lines in most jets. The method can be refined by including the
[N~I]/[N~II] line ratios integrated over the cooling zone as
we describe below.

Shock models in 1-D will not represent HH knots accurately if the cooling distances
are a substantial fraction of the size of the shock. As we shall see in 
Sec.~\ref{sec:application}, cooling distances in stellar jets are typically 10 - 100 AU. 
The width of a typical jet at $\sim$ 0.05~pc from the source is $\gtrsim$ 1000~AU, so
most HH shocks are at least an order of magnitude larger than the cooling distances. 
In several cases, spatial offsets between H$\alpha$ and [S~II] exist along the entire
extent of a well-defined shock front. Such objects are ideal for the analysis described here.
Shocks that fragment into clumps $\lesssim$ 100~AU in size will be problematic to analyze
with this method.
Time-resolved observations of some HH knots have shown significant H$\alpha$ flux
variations at the shock in response to variation in the preshock density, but the cooling zones
are more stable, with flux variations on the order of factors of two over a decade
and no strong evidence for variations in the line ratios there
\citep{hartigan11}. Because our analysis relies upon integrated fluxes over the cooling
zone, with H$\alpha$ used only to identify the location of the shock (H$\alpha$ intensity
is not used), the assumption of steady-state is a good one as long as the shocks
under consideration have had enough time to develop a full cooling zone. This cooling
time is typically a decade or two, and only a handful of HH shocks are young enough
for this to be a concern.

\subsection{Shock Models and Diagnostic Diagrams} 
\label{sec:models}

\subsubsection{Shock Model Grid}

The shock models are based on the Raymond-Cox 1-D steady-state
code \citep{raymond79}. Solar-like logarithmic elemental abundances 
were used, H: He: C: N: O: Ne: Mg: Si: S: Ar: Ca: Fe: Ni = 12.00: 10.93:
8.52: 7.96: 8.82: 7.92: 7.42: 7.52: 7.20: 6.80: 6.30: 7.60: 6.30.
The thermal and density structure of the postshock regions do
not depend strongly upon the abundances, though ratios of line
fluxes from different elements vary in tandem with their 
respective elemental abundances. 
The code follows the non-equilibrium distribution of
all ionization states of the 13 elements listed above, and includes
all the relevant atomic physics such as collisional ionization,
photoionization, autoionization, recombination and dielectronic
recombination, and charge exchange.
We have incorporated the latest atomic parameters for Fe~II given by
\citet{feii} into our calculations.
The models track the populations of each excited state of
all the ions through a matrix of Einstein-A values and collision strengths
for each transition. For example, populations in 159 levels of
Fe~II are derived in each time step after accounting for over 4000 
emission line transitions.  Radiative losses from both emission-line
and continuum processes such as 2-photon radiation and bremsstrahlung
are included in the energy equation to predict how the
fluid variables change with time. The code employs several step sizes
designed to track any regions of abrupt changes in the ionization
states of the elements. We chose these as appropriate for each of
the models we computed.

The radiative transfer is performed iteratively as follows. At the
start of the simulation, we use a
radiation field from a previous shock model with similar input parameters,
and track how this spectrum is absorbed downstream
in the postshock gas. Meanwhile, all the radiative emission lines are
stored in an output radiation file, binned in 1.0 eV intervals beginning
at 3.6 eV with 150 bins in all. At the end of the simulation the output
radiation file is used as the input for the next iteration. Typically this
procedure converges within two or three iterations to the final model,
which predicts fluxes for all the emission lines and generates temperature,
velocity, and density profiles throughout the cooling zone.

The grid consists of 8470 models. There are 11 choices for the
preshock number density of nucleons n$_\circ$ (cm$^{-3}$),
with log(n$_\circ$) = 2.0 - 4.0 in increments of 0.2.
The ten choices for preshock ionization fraction
X$_H$ = n$_\circ$(HI)/n$_\circ$(HI + HII) of H were 0.01, 0.03, 0.05,
0.10, 0.15, 0.20, 0.30, 0.40, 0.60, and 0.80. Preshock He is neutral
in all models. The shock velocity V$_S$ had
11 values, starting at 30 km$\,$s$^{-1}$ and ending with 80 km$\,$s$^{-1}$ with
increments of 5 km$\,$s$^{-1}$. Each set of (V$_S$, n$_\circ$, and X$_H$) 
used seven different preshock magnetic field strengths, designed to produce
Alfv\'enic Mach numbers M$_A$ = V$_S$/V$_A$ spaced regularly in log, with 
$\log_{10}$M$_A$ = 1/6, 1/3, 2/3, 1, 4/3, 5/3 and 2 (M$_A$ = 1.5, 2.2, 4.6, 10, 22, 46, and 100).
The Alfv\'en speed V$_A$ is defined by the equation V$_A^2$ = B$_\circ^2$/(4$\pi$n$_\circ$m),
where B$_\circ$ is the component of the preshock magnetic
field that lies perpendicular to the shock-normal, n$_\circ$ is the preshock number
density of nucleons, and m is the average mass per nucleon.
Step sizes are taken to be small enough so that the temperature changes by less than 5\%\ between steps
when T $>$ 10$^4$ K. Step sizes in the lower-temperature cooling zone (T $<$ $10^4$ K)
were chosen small enough so that the uncertainties in the relevant
line ratios introduced by the finite grid size were $<$ 1\%.
Models were terminated when the temperature fell below 2500~K.
The preshock temperature was 10$^4$~K in all models. This value is of
little consequence in most of the models because it represents a small fraction
of the kinetic energy of the flow incident upon the shock. The sound speed for mostly neutral
gas at 10$^4$~K is only $\sim$ 12 km$\,$s$^{-1}$.

\subsubsection{Cooling Zone Emission Structure and a New Diagnostic Diagram for Magnetic Fields in Shocks}

Fig.~\ref{fig:non-B-case} displays the typical structure present in a shock with a
very weak magnetic field. The gas in this 50 km$\,$s$^{-1}$ model enters the
shock with a low ionization fraction of 3\%, and reaches a postshock temperature 
of about 77000~K. Rapid cooling, due in large part to collisional excitation
of neutral H, quickly drops the temperature to $\sim$ 10$^4$~K after a distance
of 0.1~AU, with a more gradual decline that extends over a cooling zone of
$\sim$ 2~AU. The ratio of magnetic to thermal pressure rises as the gas cools,
but remains less than unity throughout, so the magnetic field has no effect
on the gas dynamics in this case. The density n rises as the temperature drops
to keep approximate pressure balance, producing a total compression of about 70
at 2~AU, where the ionization fraction n$_e$/n (inferred from the separation
of the dashed and dotted curves in the lower panel) is about 10\%. The spatial
distribution of the line emission in the upper panel shows a clear offset between
H$\alpha$, which occurs at the shock, and the forbidden lines. The average
distance of the [S~II] emission from the front is 0.72~AU.

A magnetic case is shown in Figs.~\ref{fig:B-case} and~\ref{fig:B-case-zoom}. In this model, the
preshock magnetic field is 174~$\mu$G, which results in an Alfv\'enic Mach number
of 4.6. The magnetic pressure dominates the gas pressure beyond about 0.1~AU,
so the density remains nearly constant in the cooling zone, with a compression
n/n$_o$ of about 6, and an ionization fraction of $\sim$ 10\%. 
This compression is much lower than in the weaker magnetic model in Fig.~\ref{fig:non-B-case}.
The field reduces the immediate postshock temperature slightly relative to the
nonmagnetic case, but the main difference between the two cases is that the
field extends the cooling zone size by a factor of about 20. The average [S~II]
photon now emits at 19~AU from the shock front.

The new models are in agreement with past work \citep{hmr94} which showed that emission line ratios 
integrated over the entire cooling zone
are degenerate with respect to field strength in the sense that line ratios from
a shock with a small magnetic field resemble those from a shock with a stronger field,
higher preshock density and higher shock velocity.  However, when we include
the cooling zone distance d$_C$ as a constraint, and instead of the field strength
we use the Alfv\'enic Mach number as the magnetic observable,
we obtain the remarkable result shown in Fig.~\ref{fig:rbd}. 
The figure shows that different Alfv\'enic Mach numbers separate in a
plot of the [S~II] $\lambda$6716/$\lambda$6731 line ratio integrated over the
cooling distance against d$_C$.  As both the [S~II] ratio and d$_C$ are straightforward
to measure for any resolved cooling zone, it is easy to use this diagram to
measure M$_A$ from observed spectra.

\subsubsection{The Physics Behind the New Diagnostic}

\begin{figure}%[t!] %fig4
\centering
\includegraphics[scale=1.00]{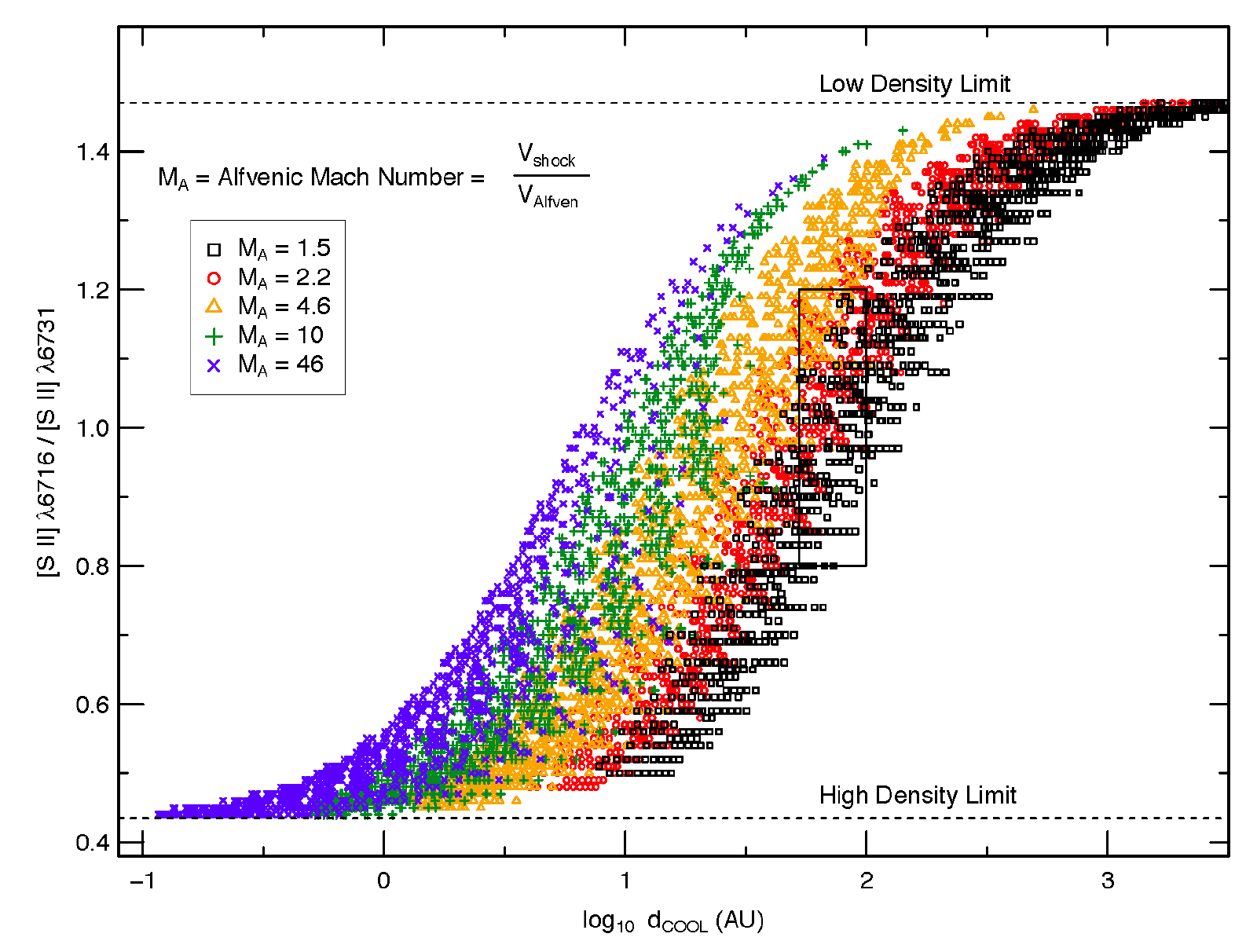}
\caption{Different Alfv\'enic Mach numbers plotted as a function of two observables, the
[S~II]$\lambda$6716/$\lambda$6731 emission line ratio and the cooling distance,
easily measured from the spatial offset between H$\alpha$ and [S~II] emission. The separation
of Alfv\'enic Mach numbers in this diagram provides a means to estimate the importance of
magnetic fields in the flow dynamics. The boxed area represents current observational uncertainties
for knot K in HH~111. 
}
\label{fig:rbd}
\end{figure}

While the results displayed in Fig.~\ref{fig:rbd} can be taken at face value as arising from
a reliable and well-tested model that includes all the relevant physics of radiative shocks,
it is a worthwhile exercise to try to understand the main physics that underlies
the reason why models with different Alfv\'enic Mach numbers separate in Fig.~\ref{fig:rbd}. 
Let us denote the preshock gas with the subscript `o', the
gas immediately behind the shock with subscript `1', and the point in the cooling
zone where the magnetic and thermal pressures are equal with subscript `2'.
The Alfv\'enic Mach number M$_A$ is defined in terms of the shock velocity V$_S$,
mass per nucleon m, preshock density n$_o$, and preshock magnetic field B$_o$ as

\begin{equation}
\label{eqn:Ma}
M_A^2 = {{4\pi n_o m V_S^2}\over{B_o^2}}.
\end{equation}

\noindent
Equating the thermal and magnetic pressure at position 2, and using the fact that
B $\sim$ n throughout the shock we obtain

\begin{equation}
\label{eqn:B2}
{{B_2^2}\over{8\pi}} = {{B_o^2n_2^2}\over{8\pi n_o^2}} = n_2kT_2.
\end{equation}

\noindent
If we now make the approximation that the pressure between points 1 and 2 is
constant, and employ the jump conditions for a strong shock we have

\begin{equation}
\label{eqn:Bo}
{{B_o^2n_2^2}\over{8\pi n_o^2}} = n_1kT_1 = {{3n_1mV_S^2}\over{16}} = {3\over 4}n_omV_S^2.
\end{equation}

\noindent
Combining equations \ref{eqn:Ma} and \ref{eqn:Bo} we get

\begin{equation}
\label{eqn:ma-compr}
M_A = \sqrt{{2\over 3}} {{n_2}\over{n_o}}.
\end{equation}

\begin{figure}%[t!] %fig1
\centering
\includegraphics[scale=0.79]{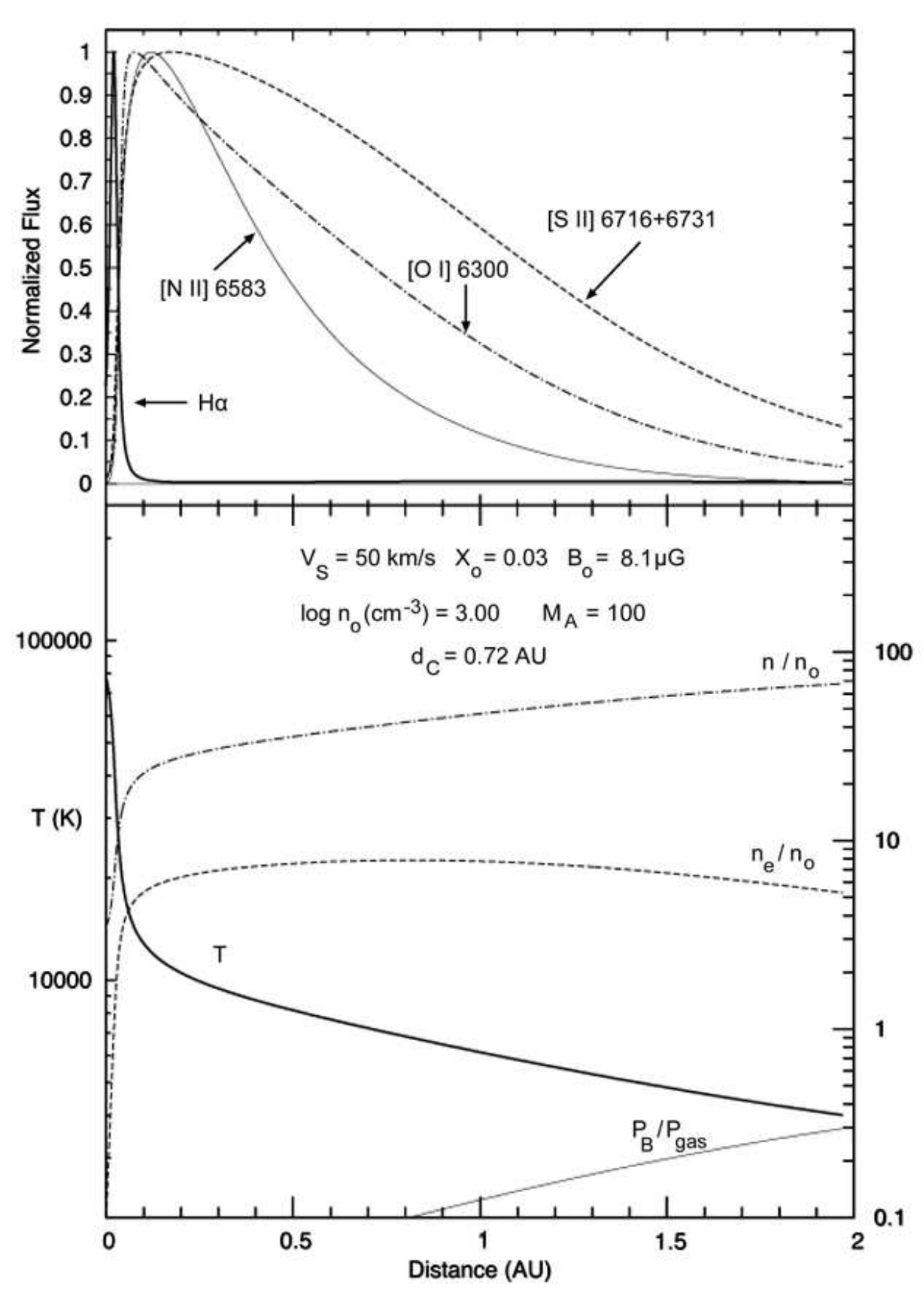}
\caption{Structure of the cooling zone behind a typical low-velocity radiative shock.
Top: Spatial distribution of radiation from several emission lines, normalized to their
peak values and plotted as a function of distance from the shock.
Bottom: Physical conditions in the postshock region. Temperature is read from the scale at
left, and the ratios n/n$_\circ$, n$_e$/n$_\circ$, and P$_B$/P$_{gas}$ refer to the scale
at the right. Here, n$_\circ$ is the preshock number density of nucleons, n and n$_e$
are the number densities of nucleons and electrons, respectively, in the postshock region, 
and P$_B$ / P$_{gas}$ is the ratio of the magnetic pressure to gas pressure. The model
parameters of V$_S$, X$_\circ$, B$_\circ$, n$_\circ$, and M$_A$ refer to the shock velocity,
preshock ionization fraction, preshock magnetic field strength, preshock density, and incident 
Alfv\'enic Mach number, respectively. The variable d$_C$ is the [S~II]-weighted cooling-zone distance, defined
in the text. This model is nonmagnetic in the 
sense that magnetic pressure is less than the gas pressure everywhere. 
}
\label{fig:non-B-case}
\end{figure}

In the magnetic-dominated regime the density remains approximately constant
because the total pressure $\sim$ B$^2$ is constant and B $\sim$ n. Hence,
equation \ref{eqn:ma-compr} implies that the compression in the forbidden-line emission zone
(relative to the preshock density) is $\sim$ 1.2 M$_A$.  For example, in Fig.~\ref{fig:non-B-case}
the compression n/n$_\circ$ $\sim$ 6 in the forbidden emission zone, with M$_A$ $\sim$ 5.

We now use these relations to help understand the points in Fig.~\ref{fig:rbd}. The y-axis of 
Fig.~\ref{fig:rbd} is the flux ratio of [S~II]$\lambda$6716/$\lambda$6731. As is
well-known from ISM physics \citep[e.g.][]{osterbrock06, hartigan08}, there are five
low-lying states for any p$^3$ electronic configuration, and the relative
populations of the $^2$D$_{5/2,3/2}$ states depend upon the electron density.
Between the low-density limit (LDL, n$_e$ $\sim$ 20 cm$^{-3}$)
and the high-density limit (HDL, n$_e$ $\sim$ $2\times10^4$ cm$^{-3}$),
I$_{\lambda 6716}$/I$_{\lambda 6731}$ scales roughly linearly with log$_{10}$(n$_e$),
with a slope of $\sim$ $-0.5$. Assuming for the moment that cooling
zones have the same ionization fractions, the y-axis in Fig.~\ref{fig:rbd} becomes

\begin{equation}
\label{eqn:yn}
y = {{I_{\lambda 6716}}\over{I_{\lambda 6731}}} \sim {\rm constant} - 0.5\,\log_{10}{n_2}
\end{equation}

We now consider the cooling distance d$_C$. Let the volume emission coefficient
be $\epsilon$(T) (erg$\,$cm$^3$s$^{-1}$). Balancing the ram energy flux incident upon the shock
(ignoring incident thermal and magnetic energies, which are small) with the
cooling we have

\begin{equation}
\label{eqn:energy}
\frac{1}{2} mn_oV_S^3 \sim \int_0^{d_C} \epsilon(T(x))n^2(x)dx \sim \bar{\epsilon}\bar{n}^2 d_C
\end{equation}

The average emission coefficient $\bar\epsilon$ is a complex function of the
temperature, and hence V$_S$, because different ionization states of different
elements dominate the cooling at distinct zones throughout the shock.
In the non-magnetic case when shock velocities exceed $\sim$ 100 km$\,$s$^{-1}$, 
numerical models show that
d$_C$ $\sim$ n$_\circ^{-1}$ V$_S^q$, where q $\sim$ 4.2 $-$ 4.6 \citep{mh80,hrh87}.
In these shocks, the postshock temperature exceeds $10^5$~K, a regime where equilibrium
cooling curves show that $\epsilon$ declines with T \citep{rs77,schure09}. Hence,
the cooling-zone volume is mostly high-temperature material, so a reasonable
estimate for the average density in the volume is one that is proportional to the 
postshock density immediately behind shock, which in turn is a factor
of four higher than the preshock density. Hence, we can estimate $\bar{n}$ to be
proportional to n$_\circ$. For $\bar{\epsilon}$ $\sim$ T$^{p/2}$ $\sim$ V$_S^{p}$,
equation~\ref{eqn:energy} implies d$_C$ $\sim$ n$_\circ^{-1}$ V$_S^{3-p}$. The numerical
simulations imply an average p $\sim$ $-$1.5 for T $\gtrsim$ $10^5$~K, 
consistent with the characteristic slope for these temperatures
in the equilibrium cooling curves.

\begin{figure}%[t!] %fig2
\centering
\includegraphics{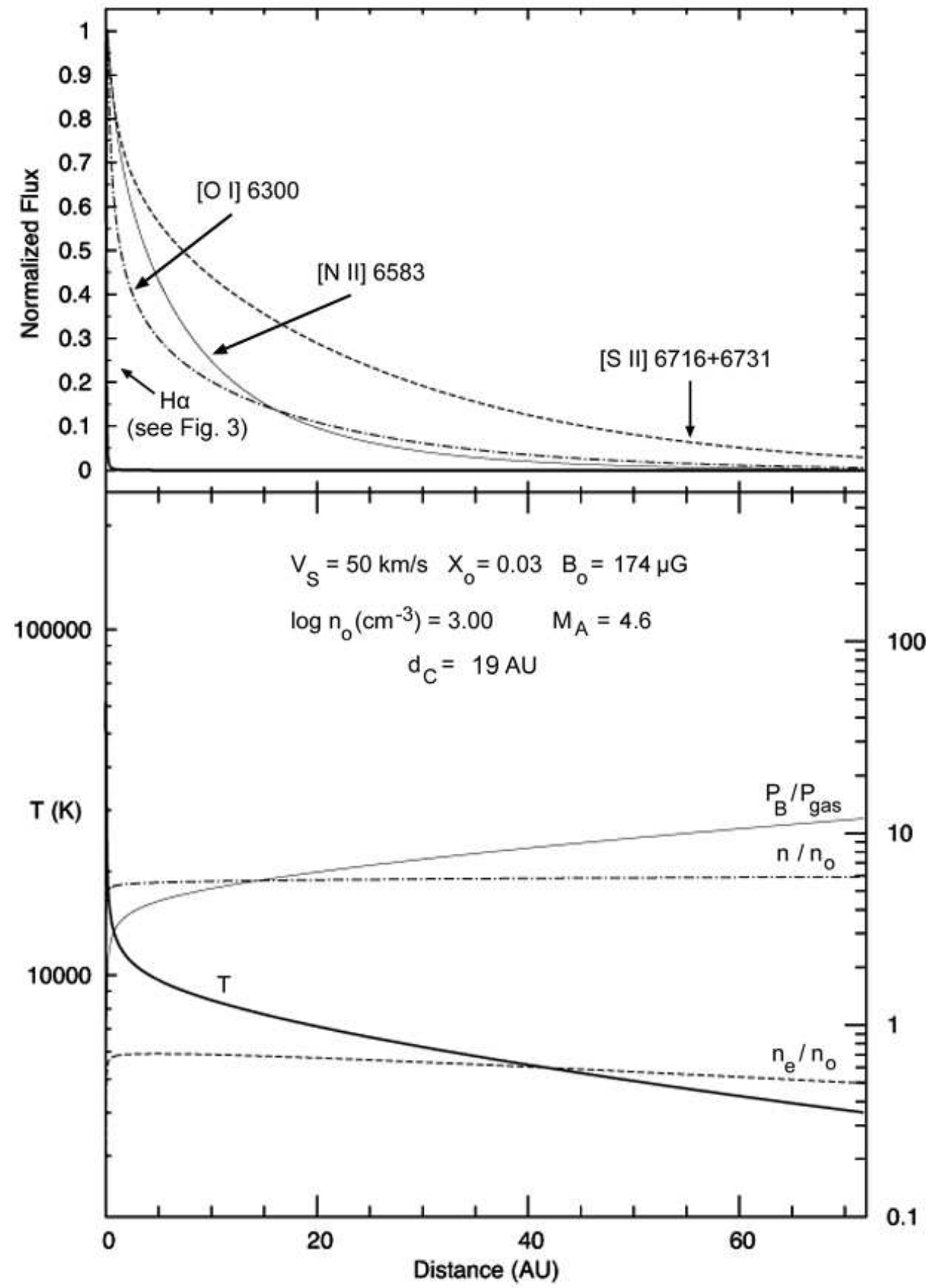}
\caption{Same as Fig.~\ref{fig:non-B-case} but for a stronger preshock magnetic field. The
magnetic field pressure expands the cooling zone distance relative to the nonmagnetic case.
}
\label{fig:B-case}
\end{figure}

\begin{figure}%[t!] %fig3
\centering
\includegraphics{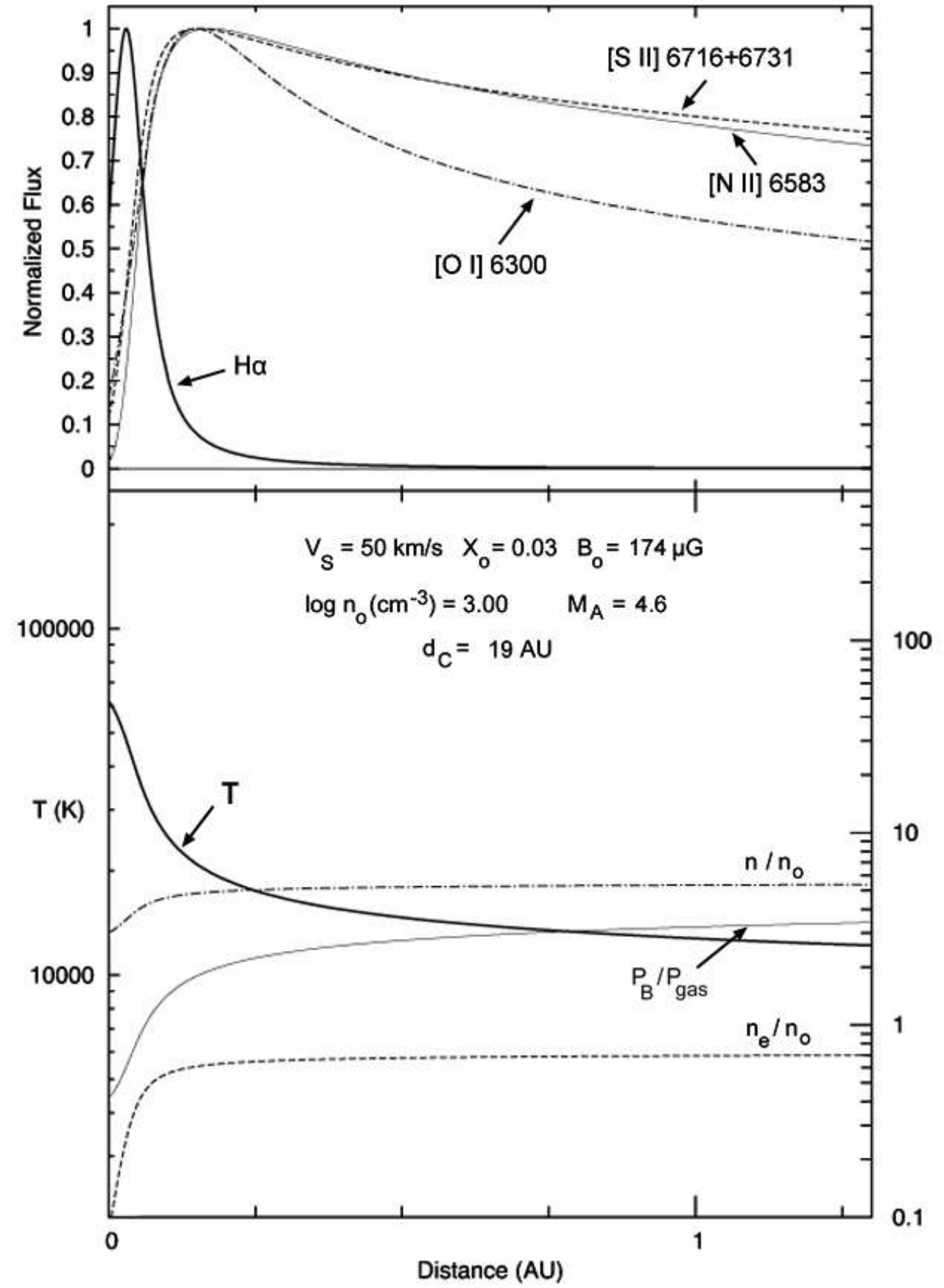}
\caption{Same as Fig.~\ref{fig:B-case} with an expanded scale on the x-axis.
}
\label{fig:B-case-zoom}
\end{figure}

The low-velocity magnetized shocks we are considering in this paper
behave differently from their high-velocity non-magnetic counterparts for two
reasons. First, the shock velocities are not high enough to fully preionize H. As a
result, gas in low-velocity shocks cools rapidly through collisional excitation of
H immediately behind the shock (Fig.~\ref{fig:non-B-case}), so that most of the
gas in the cooling zone has a temperature significantly lower than that of the 
immediate postshock gas.
Second, as shown in Figs.~\ref{fig:B-case} and \ref{fig:B-case-zoom}, 
once the magnetic pressure exceeds the thermal pressure, the gas maintains 
nearly a constant density while it cools. Hence, high-temperature
material near the shock front occupies an even smaller fraction of the total
cooling volume in the magnetized case (Fig.~\ref{fig:B-case}) than it does
for non-magnetic low-velocity shocks (Fig.~\ref{fig:non-B-case}).

The above discussion implies that for our suite of models,
the average density $\bar{n}$ in the cooling zone is approximated well by n$_2$,
not by a constant times n$_\circ$ as it would be in the non-magnetic case.
Taking $\bar{n}$ $\sim$ n$_2$ and solving for d$_C$ in equation~\ref{eqn:energy}
we get

\begin{equation}
\label{eqn:dc}
d_C \sim f(V_S)M_A^{-1}n_2^{-1}
\end{equation}

\noindent
where $f$(V$_S$) = V$_S^3$/$\bar\epsilon$ is a function of the shock velocity V$_S$, and 
we have used equation \ref{eqn:ma-compr} to relate n$_o$/n$_2$ to M$_A$.  
As above, taking $\bar{\epsilon}$ $\sim$ V$_S^p$ and
$f$(V$_S$) $\sim$ V$_S^{3-p}$, where p is an approximate power law index gives

\begin{equation}
\label{eqn:xy}
y = {\rm constant} + 0.5\log_{10}{d_C} -0.5(3-p)\log_{10}{V_S} + 0.5\log_{10}{M_A}
\end{equation}

\noindent
Hence, models with fixed V$_S$ and fixed M$_A$ but different n$_\circ$ and B$_\circ$
follow an approximately linear curve in Fig.~\ref{fig:rbd}, with the higher-density models
having a shorter cooling zones (d$_C$ $\sim$ f(V$_S$)M$_A^{-2}$n$_o^{-1}$ from
equations~\ref{eqn:ma-compr} and~\ref{eqn:dc}) and correspondingly
lower I$_{\lambda 6716}$/I$_{\lambda 6731}$ ratios.  Alternatively, if
we fix V$_S$ and d$_C$ allow both n$_o$ and M$_A$ to vary, the positive
sign in front of the $\log_{10}{M_A}$ term in equation~\ref{eqn:xy} implies
that models with larger M$_A$ trend upward in Fig.~\ref{fig:rbd}. This is the physical reason
for the color offsets in that figure.
In hindsight, that the models in Fig.~\ref{fig:rbd} separate 
when we sort by the dimensionless parameter of Alfv\'enic Mach number rather than by 
the magnetic field strength B makes sense physically because M$_A$ determines
how fast the flow is relative to the magnetic signal speed, and is
the factor that determines the compression behind the shock.

The points in Fig.~\ref{fig:rbd} also shift as the shock velocity V$_S$ and the preshock
ionization fraction change, but these effects are significantly smaller than the magnetic
effect and effectively introduce some scatter or `width' to each color band.
The shock velocity term in Eqn.~\ref{eqn:xy} has an opposite sign to that of
the M$_A$ term, and so affects the points in 
Fig.~\ref{fig:rbd} in the opposite sense, in that higher values of V$_S$ move points lower in y.
For example, the groups of 5 - 10 red points that form `strings' that
tend down and to the right in the figure are models that have the same preshock
ionization fraction, preshock density and Alfv\'enic Mach number, but increasing shock
velocities and correspondingly higher magnetic field strengths as points trend to the lower right.
Gaps between these strings of red and black points simply result from the finite number of
models run.

The magnetic offsets in Fig~\ref{fig:rbd} dominate over effects caused by
different V$_S$ because M$_A$ varies by a factor
of 67 in the grid, while V changes by less than a factor of 3.  The models indicate
that p $\sim$ 2 over the range of shock velocities in the grid, so the coefficient
in front of the log$_{10}$ V$_S$ term in equation~\ref{eqn:xy}
is essentially the same as the coefficient in front of the log$_{10}$ M$_A$ term.
Returning to the preshock ionization fraction, we find that it
also adds a small amount of scatter to Fig.~\ref{fig:rbd}. Higher preshock ionization
fractions imply higher electron densities and more cooling, which leads
to points moving down and to the left in Fig.~\ref{fig:rbd}. The effective scatter
introduced is $\sim$ 0.05 in I$_{\lambda 6716}$/I$_{\lambda 6731}$,
compared with a scatter of $\sim$ 0.08 caused by varying V$_S$. The scatter is
somewhat higher at lower M$_A$.

\subsubsection{Improving the Diagnostic Diagram}

Hundreds of possible choices exist for the y-axis in Fig.~\ref{fig:rbd}, using line ratios or
some combination of line ratios. The [S~II] $\lambda$6716/$\lambda$6731 ratio has several
advantages in that it is independent of the reddening and elemental abundances, and
is one of the brightest lines in an optical spectrum. However, as described
above, degeneracies involving V$_S$ and the preshock ionization fraction
add some scatter to the plot. The [S~II] ratio also compresses the
points near the HDL and LDL, limiting broader applicability.

We can significantly improve upon uncertainties in estimates of M$_A$
by adding additional line ratio measurements. 
Ideally, each color band in Fig \ref{fig:rbd} should be made as narrow as possible.
A new line ratio is useful in reducing the scatter in M$_A$ in Fig.~\ref{fig:rbd}
if there is a trend in a plot of [S~II] $\lambda$6716/$\lambda$6731 ratio \hbox{vs.} the new ratio
for a fixed range of d$_C$ and a given M$_A$.
Mathematically, it is the equivalent of searching for a nonzero partial derivative for a new
line ratio with respect to [S~II] $\lambda$6716/$\lambda$6731, keeping d$_C$ and M$_A$ fixed.

Keeping the constraint of bright emission lines that
don't depend upon reddening, and recognizing that we need a line ratio 
sensitive to the ionization, the [N~II]/H$\alpha$ ratio integrated over the
cooling zone emerges as a natural secondary diagnostic. 
With [S~II] already a diagnostic of density and d$_C$, it is probably not too surprising 
that an important secondary indicator of M$_A$ could come from [N~II] $\lambda$6583.
This line traces the ionization fraction of N because the N~II/N~I ratio is
tied to that of H~II/H~I through a strong charge exchange coefficient in the
plasma.  Another ratio that showed promise was [S~II] $\lambda\lambda$6716+6731/H$\alpha$.

However, the problem with [N~II]/H$\alpha$, [S~II] $\lambda\lambda$6716+6731/H$\alpha$, and 
similar ratios that involve different elements is that they depend linearly upon
the elemental abundance ratios. These ratios
are typically unknown in star-forming regions, and there is even evidence that
gas-phase abundances of some elements vary along stellar jets as dust grains are
destroyed \citep{giannini04}. Abundance ratios that use different ionization stages of the
same element avoid this problem, but unfortunately there are no bright [S~I] or [S~III] lines
to complement [S~II] in most stellar jets, and one cannot easily use [O~I]/[O~II] either because
the brightest [O~II] lines ($\lambda\lambda$3727+3729) are in the blue and subject to
large and poorly-known reddening corrections, and the red [O~II] doublet lines at $\sim$ $\lambda$7325
are typically faint and often blended with a [Ca II] line \citep[e.g.][]{bbm81}.
One promising line ratio remains: [N~I] $\lambda\lambda$5198+5200 / [N~II]$\lambda\lambda$6548+6584,
which we henceforth designate simply as `N~I/N~II'. 
These [N~I] and [N~II] lines are typically bright in HH jets, differential reddening corrections are small,
and abundance uncertainties are not an issue. 

\begin{figure}%[t!] %fig5
\centering
\includegraphics[scale=1.00]{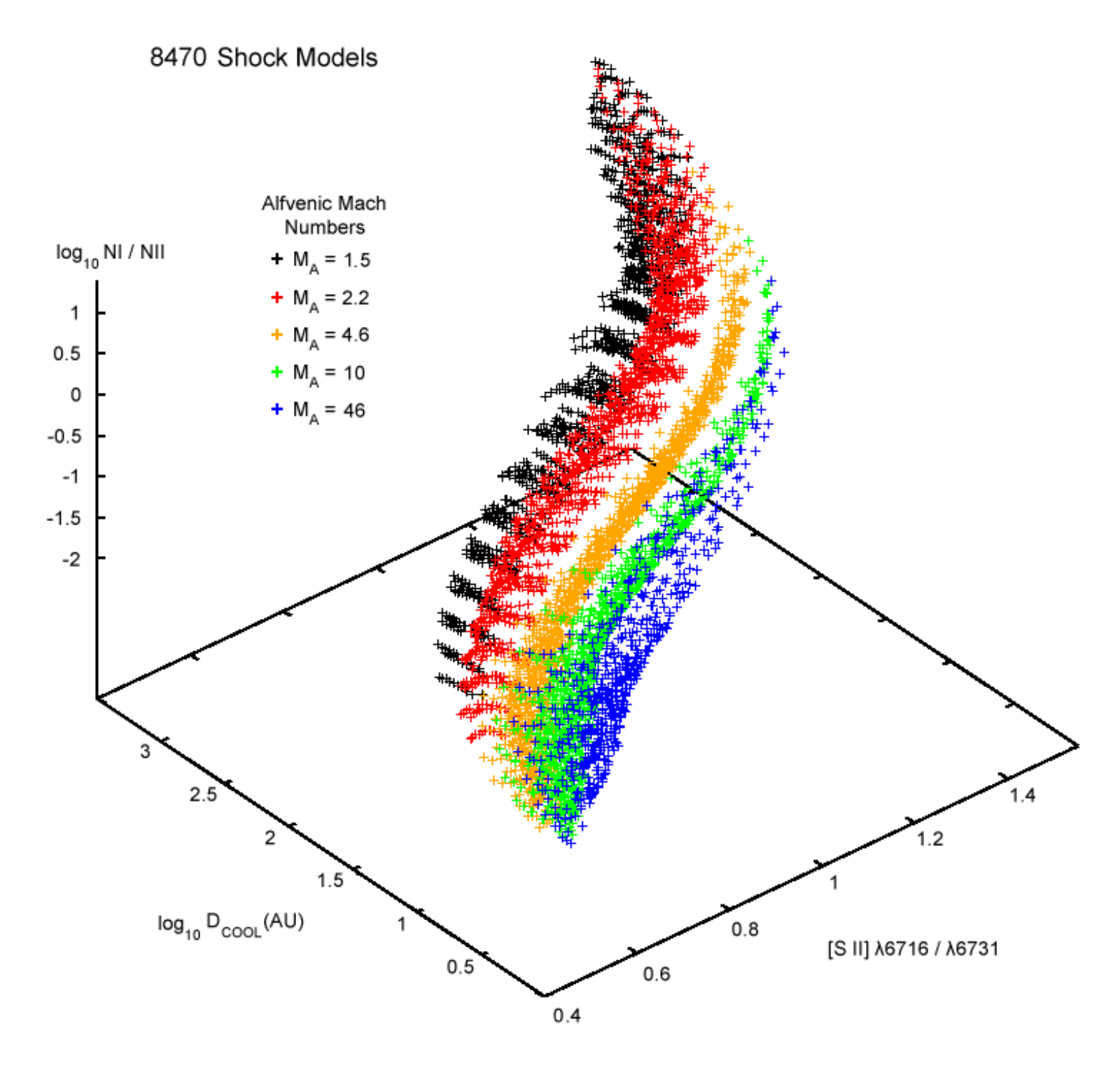}
\caption{Same as Fig.~\ref{fig:rbd} but adding in a third dimension N~I/N~II, defined as 
[N~I] $\lambda\lambda$5199+5201/[N~II] $\lambda\lambda$6548+6583.  Models with fixed
M$_A$ trace nearly parallel sheets.
}
\label{fig:3d}
\end{figure}

\begin{figure}%[t!] %fig6
\centering
\includegraphics[scale=1.00]{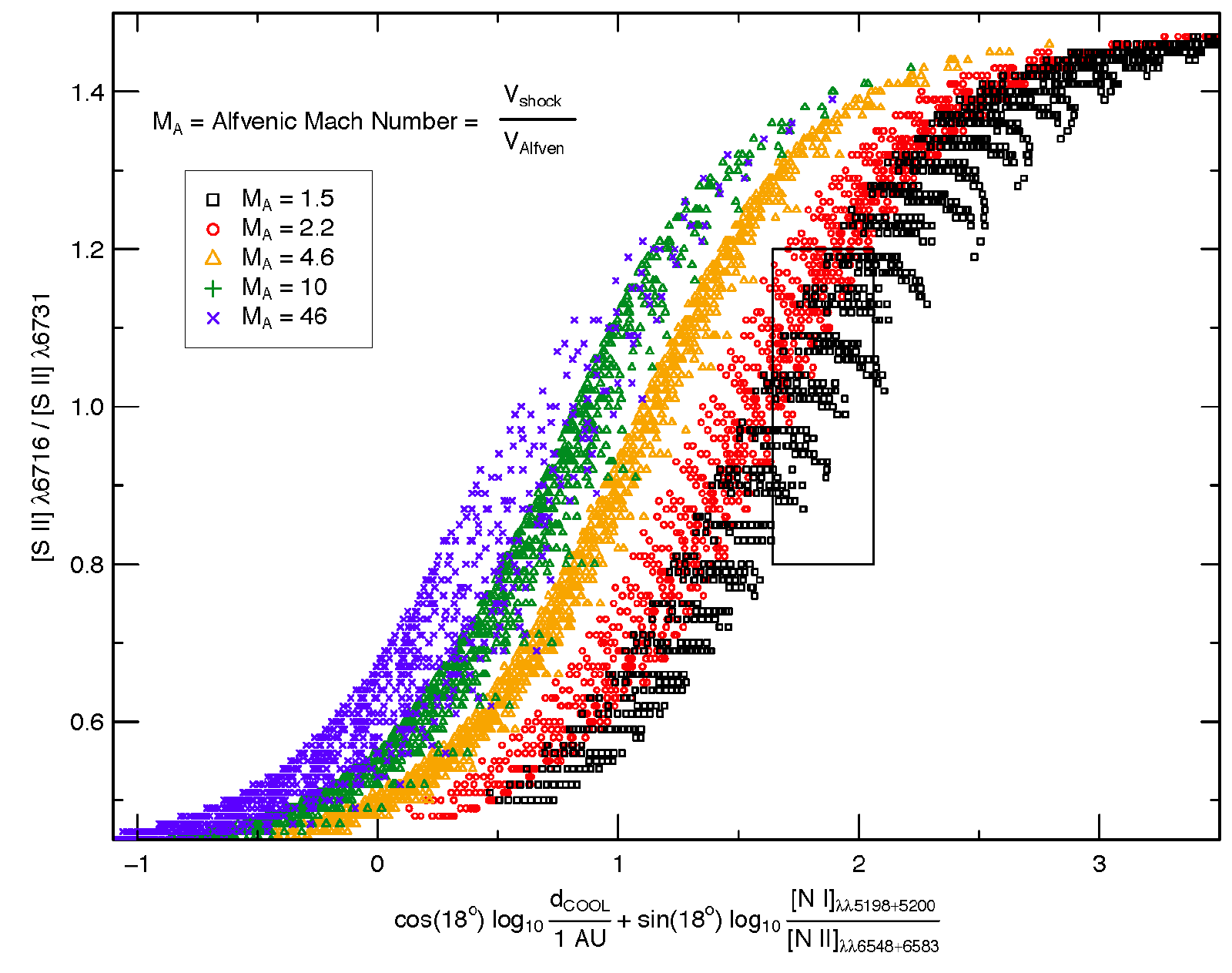}
\caption{Same as Fig.~\ref{fig:rbd} but including a correction factor to d$_C$ that is
based on the observed N~I/N~II ratio. This graph is the projection of Fig.~\ref{fig:3d} along a
line of sight oriented 18$^\circ$ from the N~I/N~II axis. The boxed region denotes the possible
range of values related to knot K in HH~111. 
}
\label{fig:finalplot}
\end{figure}

We can visualize the the constraints of additional line ratios
geometrically by plotting them along a third dimension. In this case we 
rotate the cube of
($\log_{10}$d$_C$, [S~II] $\lambda$6716/$\lambda$6731, and N~I/N~II) values in such a way as
to achieve maximum separation between surfaces of constant M$_A$.  Fig.~\ref{fig:3d} 
shows that models with constant M$_A$ define a series of nested curved sheets that 
become degenerate only in the region where d$_C$ $\lesssim$ 10~AU, a regime unresolvable
spatially with existing instrumentation in any case. There are several orientations that separate
the different M$_A$, but a particularly convenient one is when the line of sight is
in the plane defined by 
($\log_{10}$d$_C$, N~I/N~II) angled about 18 degrees from the N~I/N~II axis.
This projection, which has 
[S~II] $\lambda$6716/$\lambda$6731 along the x-axis, and a linear combination of
$\log_{10}$d$_C$ and N~I/N~II along the y-axis, is shown in Fig.~\ref{fig:finalplot}.
The width of the bands defining the red and orange points (M$_A$ = 2.2 and 4.6, respectively)
narrows by about a factor of two compared with Fig.~\ref{fig:rbd}. Most of the remaining
scatter in each color arises because surfaces of fixed M$_A$ are not planar sheets, but
curve slightly at the ends (Fig.~\ref{fig:3d}).

\begin{figure}%[t!] %fig7
\centering
\includegraphics[scale=0.78]{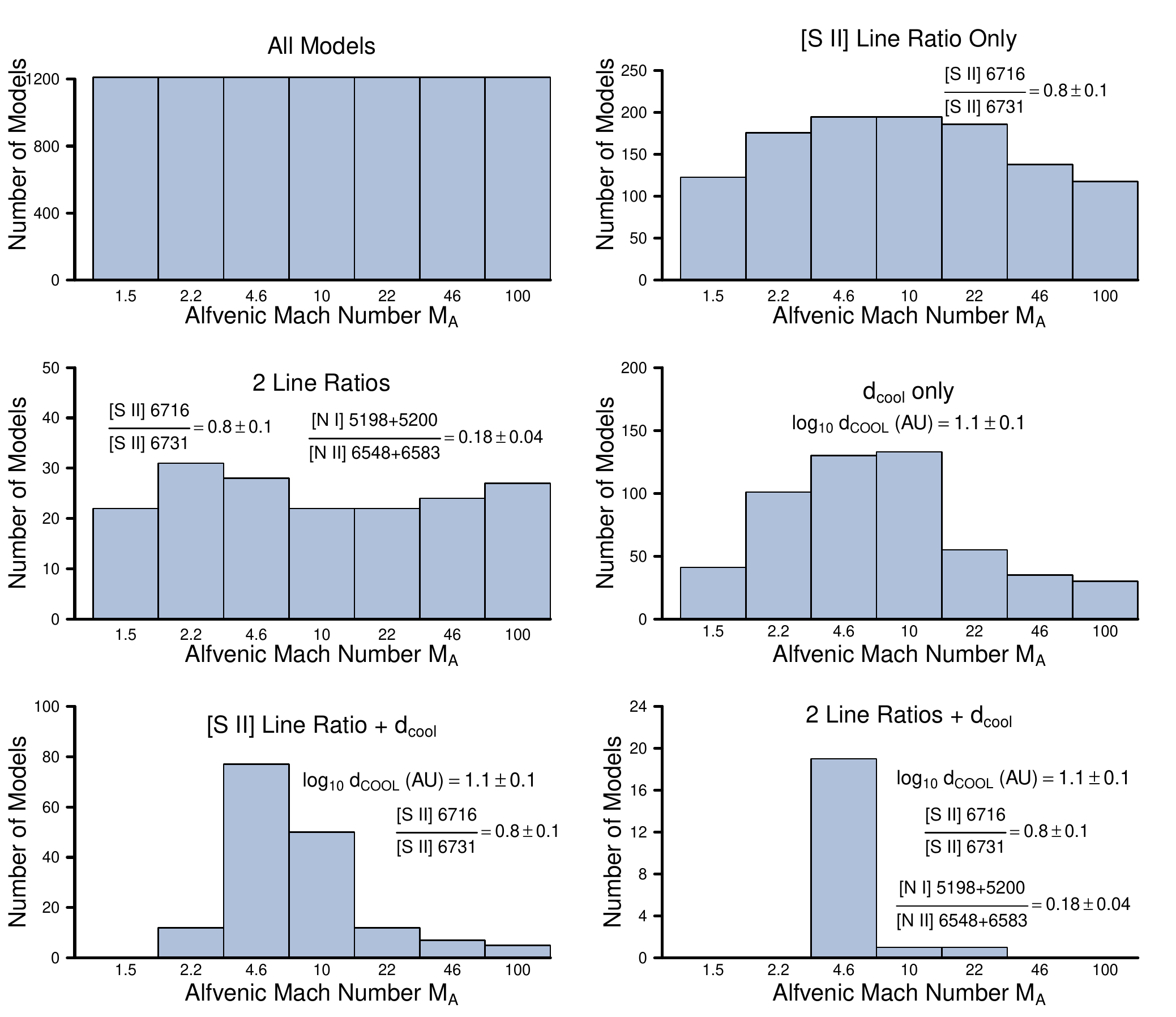}
\caption{Histograms of the number of models satisfying the listed constraint for
a hypothetical object. Alfv\'enic Mach
numbers require both line-ratio measurements and an estimate of the cooling distance
d$_{COOL}$.
}
\label{fig:histogram}
\end{figure}

Fig.~\ref{fig:finalplot} provides a quick way to plot a given observation
on a graph and determine M$_A$, but given a table of models one can also simply collect
all the models that are consistent with the data. As an example, Fig.~\ref{fig:histogram}
analyzes a hypothetical object with three observables, d$_C$, 
[S~II] $\lambda$6716/$\lambda$6731, and N~I/N~II. A total of 1131 out of the 8470 models are consistent
with the constraint [S~II] $\lambda$6716/$\lambda$6731 = 0.8 $\pm$ 0.1. That number drops
to 176 models once the constraint of [N~I]/[N~II] = 0.18 $\pm$ 0.04 is included, but the
consistent models remain spread out over a range of M$_A$. The cooling-zone observation of
$\log_{10}$ d$_{C}$ = 1.1 $\pm$ 0.1 on its own limits the number of applicable models to
525, but these are again spread out over a range of M$_A$. It is only when the
measurement of d$_C$ is combined with the [S~II] ratio (163 models) that a trend in M$_A$
emerges, and the trend becomes a narrow range of values (21 models) once both [N~II]/[N~I] and
the [S~II] ratio are combined with d$_C$. One drawback with such histograms is that the number
of models consistent with a given set of constraints depends upon the grid used to generate the models. 
Nevertheless, the peak value of the histogram always corresponds to the M$_A$ indicated
by the color plot in Fig.~\ref{fig:finalplot}.

\section{Application to Stellar Jets}
\label{sec:application}

While several examples of resolved cooling-zone distances have been
observed with HST, images of the ratio of [S~II] $\lambda$6716/$\lambda$6731 are
rare, and none seem to be available for the objects like HH~111 that have relatively
simple, well-defined shocks that move in the plane of the sky (which minimizes projection
effects). Moreover, [N~II] images are rare, and [N~I] $\lambda\lambda$5198+5201 images 
even more rare, despite the fact that both sets of lines are easily observed in long-slit spectra.

For the purposes of this paper, we choose the isolated bow shock `K' in HH~111 for further analysis
\citep{reipurth97}. This object has a resolved cooling zone of $\log_{10}$ d$_C$ = 1.86 $\pm$ 0.14
from archival HST images.  %3.5 +- 1.0 pixels = 175 +- 50 mas = 72 +- 20 AU  log10(dc) = 1.86 +- 0.14
Unfortunately, we must attempt
to recover line ratios from a mixture of ground-based and space-based spectra. 
Long-slit STIS spectra smoothed over 0.4$^{\prime\prime}$ 
published by \citet{gg03} indicate an electron density of
9000 cm$^{-3}$, but the S/N of the line ratio from these data is very low. From the ground, \citet{morse93b}
inferred N$_e$ $\sim$ 600 cm$^{-3}$ near knot K, which corresponds to  
[S~II] $\lambda$6716/$\lambda$6731 $\sim$ 1.0. However, the knot is blended to some degree
with knot L, where [S~II] $\lambda$6716/$\lambda$6731 $\sim$ 1.20. 
We adopt 1.0 for [S~II] $\lambda$6716/$\lambda$6731 in knot K, but the uncertainty is rather
large, $\sim$ 0.20 for this ratio. Taking the \citet{morse93b}
values integrated over knots D through J, the observed ratio of [N~I] $\lambda\lambda$5198+5200 /
[N~II] $\lambda$6583 is 1.30, and the reddening-corrected ratio is 2.49. From the
ratio of the A-values, the
[N~II] $\lambda$6548 flux should equal 1/3 that of [N~II] $\lambda$6583. Hence,
$\log_{10}$ (N~I/N~II) = $-$0.01 without reddening corrections, and 0.27 with reddening included.
If we use the observations for knot~L, then the reddening-corrected value of
$\log_{10}$ (N~I/N~II) = 0.15.  The [N~I]/H$\beta$ flux ratio for knot L published by \citet{morse93b}
is in agreement with that of \citet{nc93}, though we cannot use the latter reference to 
estimate [N~II] fluxes because those spectra are only at blue wavelengths.
We adopt $\log_{10}$ (N~I/N~II) = 0.27 $\pm$ 0.25 for knot K. 

The allowed ranges for the above values define the boxed
regions in Figs.~\ref{fig:rbd} and \ref{fig:finalplot}.
%log10(Dc) = 1.72-2.00; SII 6716/6731 = 0.8-1.2; cos18*log10(Dc)+sin18*log10(NI/NII) = 1.64-2.06
Despite the large uncertainties (which can be reduced significantly with 
dedicated observations), it is clear from both figures that
the Alfv\'enic Mach number will lie along the lower boundary
of the models, with M$_A$ $\sim$ 1.5 $-$ 2.2.
A total of 58 models yield values for d$_C$, N~I/N~II and [S~II] $\lambda$6716/$\lambda$6731 consistent
with the observational constraints, 25 with M$_A$ = 2.2 and 33 with M$_A$ = 1.5. 
Hence, the magnetic field is strong enough to play an important role in the dynamics of knot K in this jet.
Shock velocities, log(n$_\circ$), B$_\circ$, and X$_H$ are all poorly-constrained given the
relatively large observational errorbars.

\section{Summary}

The main point of this paper is to show that a key observable - the
cooling zone distance defined by the separation of H$\alpha$ from forbidden line
emission - makes it possible to infer Alfv\'enic Mach numbers (defined as the
shock velocity divided by the preshock Alfv\'en speed) in low-velocity shock fronts
that cool radiatively.  The physics behind why the method works is straightforward
and the effect is unavoidable: even a weak preshock magnetic field is compressed at
the shock front and continues to rise behind the front as the gas cools and the
density increases.  Once the magnetic pressure becomes comparable to the thermal pressure,
the cooling zone lengthens.

Observational requirements include being able to 
measure the spatial offset between the shock and its cooling zone, for example, from a spatial offset between
H$\alpha$ and [S~II], and a measure of the red [S~II] emission line ratio
integrated over the forbidden-line emitting zone. If the fluxes of [N~I] $\lambda\lambda$5200 doublet
and [N~II] $\lambda$6583 line are available, the models yield increasingly more precise results.
The ideal dataset has not been acquired for this type of analysis, but initial application
of the procedure to a bow shock in the HH~111 stellar jet indicates a rather low
Alfv\'enic Mach number of $\sim$ 2.  This Alfv\'enic Mach number is relative to the shock
velocities in the jet. The Alfv\'enic Mach number relative to the bulk flow speed will
be about an order of magnitude higher.

\acknowledgements
This research is supported by the Department of Energy
National Laser Users Facility (NLUF) grant DE-NA0002037. 

\clearpage

\null


\begin{thebibliography}{}

\bibitem[Bautista et~al.(2015)]{feii} 
Bautista, M.A., Fivet, V., Ballance, C., Quinet, P., Ferland, G., Mendoza, C., \& Kallman, T.R.
2015, ApJ 808, 174

\bibitem[Brugel et~al.(1981)]{bbm81} 
Brugel, E.W., B\"ohm, K.-H., \& Mannery, E. 1981, ApJS 47, 117

\bibitem[Carrasco-Gonzalez et~al.(2010)]{cg10} 
Carrasco-Gonzalez, C., Rodriguez, L.F., Anglada, G., Marti, J., Torrelles, J.M.  \& Osorio, M.
2010, Science 330, 1290

\bibitem[Chevalier et~al.(1980)]{chev80} 
Chevalier, R., Kirshner, R.P., \& Raymond, J.C. 1980, ApJ 235, 186

\bibitem[Chrysostomou et~al.(2007)]{chrys07} 
Chrysostomou, A., Lucas, P. W., \& Hough, J. H.  2007, Nature, 450, 71

\bibitem[Cox \& Raymond(1985)]{cr85} 
Cox, D.P., \& Raymond, J.C. 1985, ApJ 298, 651

\bibitem[Draine(1993)]{shockref} 
Draine, B.T. 1993, ARA\&A 31, 373

\bibitem[Frank et~al.(2014)]{frank14} Frank, A., et~al. in {\it Protostars and Planets VI}, 
H. Beuther, R. Klessen, C. Dullemond \& T. Henning eds.  (Tucson:Univ. of Arizona Press)

\bibitem[Ghavamian et~al.(2001)]{ghav01} 
Ghavamian, P., Raymond, J., Smith, R.C., \& Hartigan, P. 2001, ApJ 547, 995

\bibitem[Giannini et~al.(2004)]{giannini04} 
Giannini, T., McCoey, C., Caratti o Garatti, A., Nisini, B.,
Lorenzetti, D., \& Flower, D.R. 2004, A\&A 419, 999

\bibitem[Gonzalez-Gomez \& Raga(2003)]{gg03} 
Gonzalez-Gomez, D.I., \& Raga, A.C. 2003, RevMexAA Serie de Conf. 15, 137

\bibitem[Gurnett \& Bhattacharjee(2005)]{gb05} 
Gurnett, D.A., \& Bhattacharjee, A. 2005, {\it Introduction to Plasma Physics:
With Space Applications}, (Cambridge: Cambridge University Press)

\bibitem[Hartigan(2008)]{hartigan08} 
Hartigan, P. 2008, Lecture Notes in Physics 742, 15-42

\bibitem[Hartigan et~al.(2015)]{hartigan15}
Hartigan, P., Foster, J.~M., Yirak, K., Liao, A.S., Graham, P., Frank, A., Wilde, B.,
Blue, B., Martinez, D., Rosen, P., Farley, D., \& Paguio, R.  2015, in preparation

\bibitem[Hartigan et~al.(2011)]{hartigan11}
Hartigan, P., Frank, A., Foster, J.~M., Wilde, B., Douglas, M., Rosen, P., Coker, R.,
Blue, B., \& Hansen, F.  2011, \apj, 736, 29

\bibitem[Hartigan et~al.(1994)]{hmr94} 
Hartigan, P., Morse, J.A., \& Raymond, J.C. 1994, ApJ 436, 125

\bibitem[Hartigan et~al.(2001)]{hartigan01} 
Hartigan, P., Morse, J., Reipurth, B., Heathcote, S. \& Bally, J. 2001, ApJ 559, L157

\bibitem[Hartigan et~al.(2007)]{hartigan07} 
Hartigan, P., Frank, A., Varniere, P., \& Blackman, E. 2007, ApJ 661, 910

\bibitem[Hartigan \& Morse(2007)]{hm07} 
Hartigan, P., \& Morse, J. 2007, ApJ 660, 426

\bibitem[Hartigan et~al.(1987)]{hrh87} 
Hartigan, P., Raymond, J.C., \& Hartmann, L. 1987, ApJ 316, 323

\bibitem[Heathcote et~al.(1996)]{heathcote96} 
Heathcote, S., Morse, J., Hartigan, P., Reipurth, B., Schwartz, R.D., Bally, J., \& Stone, J.
1996, AJ 112, 1141

\bibitem[Lazendic et~al.(2006)]{laz06} 
Lazendic, J.S., Dewey, D., Schulz, N.S., \& Canizares, C.R. 2006, ApJ 651, 250

\bibitem[McKee \& Hollenbach(1980)]{mh80} 
McKee, C.F., \& Hollenbach, D.J. 1980, ARA\&A 18, 219

\bibitem[Morse et~al.(1992)]{morse92} 
Morse, J., Hartigan, P., Cecil, G., Raymond, J., \& Heathcote, S.
1992, ApJ 399, 231

\bibitem[Morse et~al.(1993)]{morse93b} 
Morse, J., Heathcote, S., Hartigan, P., \& Cecil, G.
1993b, AJ 106, 1139

\bibitem[Morse et~al.(1993)]{morse93a} 
Morse, J., Heathcote, S., Cecil, G., Hartigan, P., \& Raymond, J.
1993a, ApJ 410, 764

\bibitem[Moser \& Bellan(2012)]{moser12} 
Moser, A.L., \& Bellan, P.M. 2012, Nature 482, 379

\bibitem[Noriega-Crespo et~al.(1993)]{nc93} 
Noriega-Crespo, A., Garnavich, P.M., \& Raga, A.C. 1993, AJ 106, 1133

\bibitem[Osterbrock \& Ferland(2006)]{osterbrock06} 
Osterbrock, D.E., \& Ferland, G.J. 2005, {\it Astrophysics of Gaseous Nebulae and
Active Galactic Nuclei}, (Sausalito: University Science Books)

\bibitem[Raga \& Binette(1991)]{rb91} 
Raga, A.C., \& Binette, L. 1991, RMxA\&A 22, 265

\bibitem[Raga et~al.(2015)]{raga15} 
Raga, A.C., Reipurth, B., Castellanos-Ramirez, A., Chiang, H.-F., \& Bally, J.
2015, ApJ 798, L1

\bibitem[Reipurth et~al.(1997)]{reipurth97} 
Reipurth, B., Hartigan, P., Heathcote, S., Morse, J., \& Bally, J. 1997, AJ 114, 757

\bibitem[Ray et~al.(1997)]{ray97} 
Ray, T.P., Muxlow, T.W.B., Axon, D.J. et~al. 1997, Nature 385, 415

\bibitem[Raymond (1979)]{raymond79} 
Raymond, J.C. 1979, ApJS 39, 1

\bibitem[Raymond \& Smith(1977)]{rs77} 
Raymond, J.C. \& Smith, B.W. 1977, ApJS 35, 419

\bibitem[Raymond et~al.(2008)]{snrref} 
Raymond, J.C., Isenberg, P.A., \& Laming, J.M. 2008, ApJ 682, 408

\bibitem[Raymond et~al.(2011)]{raymond11} 
Raymond, J.C., Vink, J., Helder, E.A., \& de Laat, A. 2011, ApJ 731, L14

\bibitem[Reynolds(2008)]{reynolds08} 
Reynolds, S.P. 2008, Ann.Rev.Astr.Ap. 46, 89

\bibitem[Riera et~al.(2003)]{riera03}
Riera, A., Garcia-Lario, P., Manchado, A., Bobrowsky, M., \& Estalella, R.
2003, A\&A 401, 1039

\bibitem[Schneider et~al.(2011)]{xray}
Schneider, P.C., G\"unther, H.M., \& Schmitt, J.H.M.M. 2011, A\&A 530, 123

\bibitem[Schure et~al.(2009)]{schure09}
Schure, K.M., Kosenko, D., Kaastra, J.S., Keppens, R., \& Vink, J. 2009, A\&A 508, 751

\bibitem[Staff et~al.(2015)]{staff15}
Staff, J.E., Koning, N., Ouyed, R., Thompson, A., \& Pudritz, R.E.
2015, MNRAS 446, 3975

\bibitem[Williams(2013)]{williams13} 
Williams, R. 2013, AJ 146, 55

\end{thebibliography}
\end{document}